\documentclass[aip,apl,twocolumn,amssymb,amsmath,reprint]{revtex4-1}

\usepackage{amsfonts}
\usepackage{mathrsfs}
\usepackage{amsmath}% needed for subequations
\usepackage{color}
\usepackage{graphicx}
\usepackage{bm}% bold maths
\usepackage{amssymb}
\usepackage{xspace}
\usepackage{epstopdf}
\usepackage{dcolumn}% Align table columns on decimal point
\usepackage{longtable}
\usepackage{multirow}
\usepackage[colorlinks=true, letterpaper=true, pdfstartview=FitV, linkcolor=blue, citecolor=blue, urlcolor=blue]{hyperref}

\begin{document}

\title{The centrosymmetric Li$_2$NaN material: an ideal topological semimetal with critical-type triply degenerate nodal points}

\author{Lei Jin}
\affiliation{School of Materials Science and Engineering, Hebei University of Technology, Tianjin 300130, China.}

\author{Xiaoming Zhang}
\email{zhangxiaoming87@hebut.edu.cn}
\affiliation{School of Materials Science and Engineering, Hebei University of Technology, Tianjin 300130, China.}

\author{Xuefang Dai}
\affiliation{School of Materials Science and Engineering, Hebei University of Technology, Tianjin 300130, China.}

\author{Heyan Liu}
\affiliation{School of Materials Science and Engineering, Hebei University of Technology, Tianjin 300130, China.}

\author{Guifeng Chen}
\affiliation{School of Materials Science and Engineering, Hebei University of Technology, Tianjin 300130, China.}

\author{Guodong Liu}
\email{gdliu1978@126.com}
\affiliation{School of Materials Science and Engineering, Hebei University of Technology, Tianjin 300130, China.}

\begin{abstract}
Triply degenerate band crossing can generate novel triply degenerate nodal point (TDNP) fermions. For the TDNP character to the best manifest in physical properties, it desires that the TDNPs are near the Fermi level, far away with each other in the reciprocal space, and not coexisting with other quasiparticles or extraneous bands. Here we predict that the centrosymmetic material Li$_2$NaN is a realistic TDNP semimetal that meets all these requirements. Li$_2$NaN features only a single pair of TDNPs at the Fermi level, under the protection of the \emph{C$_{6v}$} symmetry. Interestingly, the TDNPs identified here show the critical-type band crossing, which is different from those in previously reported TDNP semimetals. The Fermi arc surface states for the TDNPs have been identified. Under lattice strain, we show that the position of the TDNPs can be tuned, and the TDNPs can even be annihilated. The centrosymmetry in Li$_2$NaN makes the TDNP transform into a critical type Dirac point under the spin-orbit coupling, which is drastically distinct from those in noncentrosymmetric systems. Such centrosymmetic TDNP semimetal has been rarely identified in realistic materials.
\end{abstract}

\maketitle

Materials with nontrivial band topology have attracted great interests in recent years. Beside their potential technological applications, topological materials provide the possibility to study the fundamental physics theories in the relatively convenient condensed matter scale. For example, Weyl (Dirac) semimetals~\cite{1,2,3,4,5,6,7,8,9,10,11,12,13,14} host Weyl (Dirac) points with doubly (fourfold-) degenerate linear band crossings near the Fermi level, and their low-energy excitations can serve as the analogues of Weyl (Dirac) electrons in high-energy theories. More interestingly, three types of quasiparticle excitations, namely triply, six, and eightfold-degenerate nodal points, which even do not have high-energy counterparts, have also been proposed in realistic topological materials~\cite{15,16}. In particular, the triply degenerate nodal points (TDNPs)~\cite{17,18,19,20,21,22,23,24,25,26}, which are formed by the crossing between a non-degenerate band and a doubly-degenerate band, are proposed to host intriguing physics, such as large negative magnetoresistance~\cite{27}, helical anomaly~\cite{20}, exotic Fermi surface transitions~\cite{19,20}, and unconventional quantum Hall effects~\cite{28}.

Initially, TDNP semimetals are predicted in some inversion-asymmetric materials, including the strained-HgTe~\cite{17}, WC-type materials~\cite{19,20,21,22}, Heusler compounds~\cite{24,25}, NaCu$_3$Te$_2$ family~\cite{26}, PtBi$_2$ compound~\cite{29}, etc. Recently, Zhang \emph{et al}.~\cite{23} first reported the existence of TDNPs in the centrosymmetric TiB$_2$ materials by theoretical prediction. They found the TDNPs in these centrosymmetric materials can transform into a novel type of Dirac points, which are drastically different from those in inversion-asymmetric systems. Beside theoretical predictions, several angle-resolved photoemission spectroscopy (ARPES) and transport experiments have also been carried out to probe the electronic structures around TDNPs in WC-type materials~\cite{27,30,31,32,33} and the trigonal layered PtBi$_2$~\cite{29}. However, the experimental detection on the intrinsic properties of TDNP fermions is still in challenge, mainly due to the lack of ideal candidate materials. For an ideal TDNP semimetal, candidate materials should at least meet the following requirements. First, the TDNPs should appear near the Fermi level. Second, it is the best that the materials only possess a single pair of TDNPs. Third, the TDNPs are far away with each other in the reciprocal space. Fourth, the TDNPs do not coexist with the other quasiparticles or extraneous bands. These rigorous requirements have led to the great scarcity of suitable TDNP candidate materials. Therefore, to facilitate the experimental investigation on the properties of TDNP semimetals, it is in urgent need to search for ideal candidate materials that meet these requirements.

In the present work, based on the first-principles calculations, we report that the centrosymmetric Li$_2$NaN is an ideal TDNP semimetal which meets all the requirements mentioned above. More interestingly, the TDNPs in Li$_2$NaN are formed by the crossing of a flat band (double-degenerate) and a dispersive band (non-degenerate). Such critical-type TDNPs with unique slope of band crossing are different from previously studied ones. The nontrivial surface states from TDNPs are revealed. Moreover, we find that strain can effectively tune the position of the TDNPs and even annihilate the TDNPs to drive the system into a nodal line phase. Because of the presence of inversion symmetry, the TDNPs transform into critical-type Dirac points under spin-orbit coupling (SOC). Considering the SOC effect is negligible in the Li$_2$NaN material, the TDNP signatures should clearly manifest for experimental detections. Our work offers an ideal platform for experimentally investigating the novel properties of TDNP semimetals.

We perform first-principles calculations in the framework of density functional theory using the projector augmented wave method, as implemented in the Vienna ab initio Simulation Package~\cite{34,35,36}. For ionic potentials, we use the generalized gradient approximation with the Perdew-Burke-Ernzerhof realization~\cite{37}. The cutoff energy is chosen as 500 eV, and the Brillouin zone (BZ) is sampled with a $\Gamma$-centered $k$-mesh of size 13$\times$13$\times$13 for geometry optimization and self-consistent calculations. The energy convergence criterion is set as $10^{-6}$ eV. To investigate the topological surface states, we construct the localized Wannier functions~\cite{38,39}, and calculate surface spectra by using the WANNIERTOOLS package~\cite{40} combined with the iterative Green¡¯s function method~\cite{41}.

\begin{figure}
%\scalebox{0.4}{\includegraphics*{Figure1.pdf}}
\includegraphics[width=8cm]{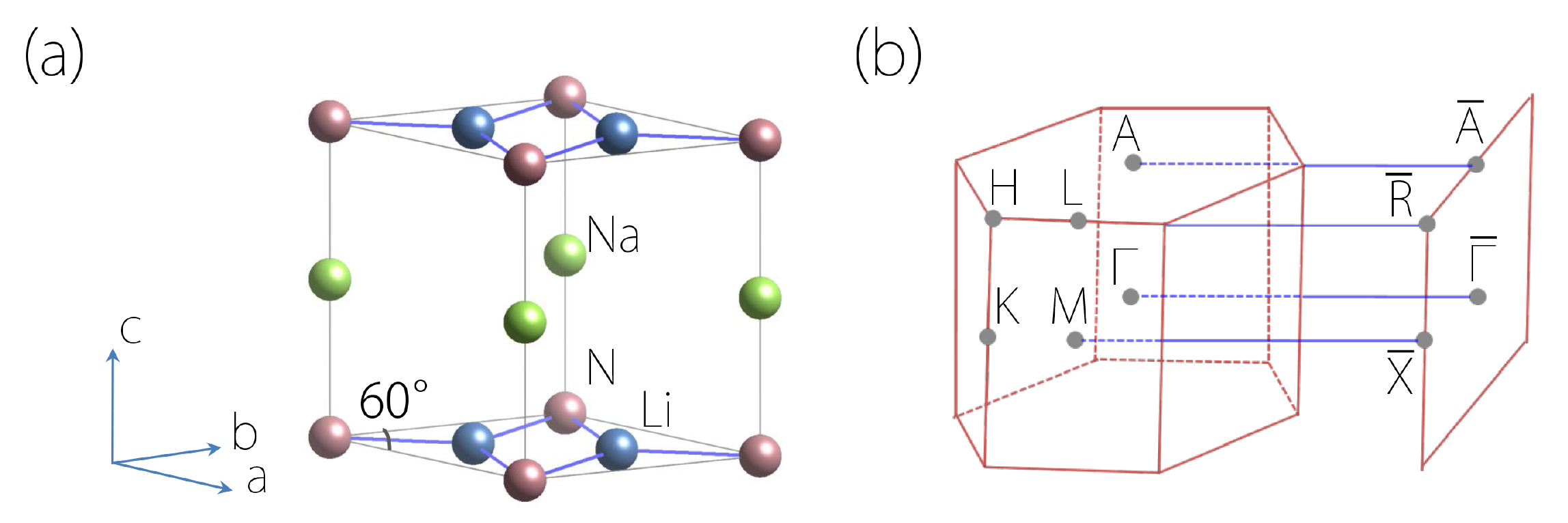}
\caption{(a) Crystal structure of the Li$_2$NaN compound. (b) The bulk Brillouin zone and its projection onto the (010) surface. The high-symmetry points are labelled.}
\label{fig1}
\end{figure}

The Li$_2$NaN material (ICSD No.: 92308) has a centrosymmetric hexagonal crystal structure with the space group \emph{P6/mmm }(No. 191). As shown in Fig.~\ref{fig1}(a), the bonding of Li and N atoms forms a coplanar Li-N layer and stacks with the Na layer along the c axis. In the unit cell, Li, Na and N atoms occupy the \emph{2c} (1/3, 2/3, 0), \emph{1b} (0, 0, 1/2) and \emph{1a} (0, 0, 0) Wyckoff sites, respectively. The fully relaxed lattice constants are \emph{a=b}=3.658\AA ~, \emph{c}=4.761\AA ~, which match well with the experimental values (\emph{a=b}=3.65\AA ~ and \emph{c}=4.60\AA ~)~\cite{42}. The optimized lattice structure is used in our calculations.

\begin{figure}
%\scalebox{0.2}{\includegraphics*{Figure2.pdf}}
\includegraphics[width=8cm]{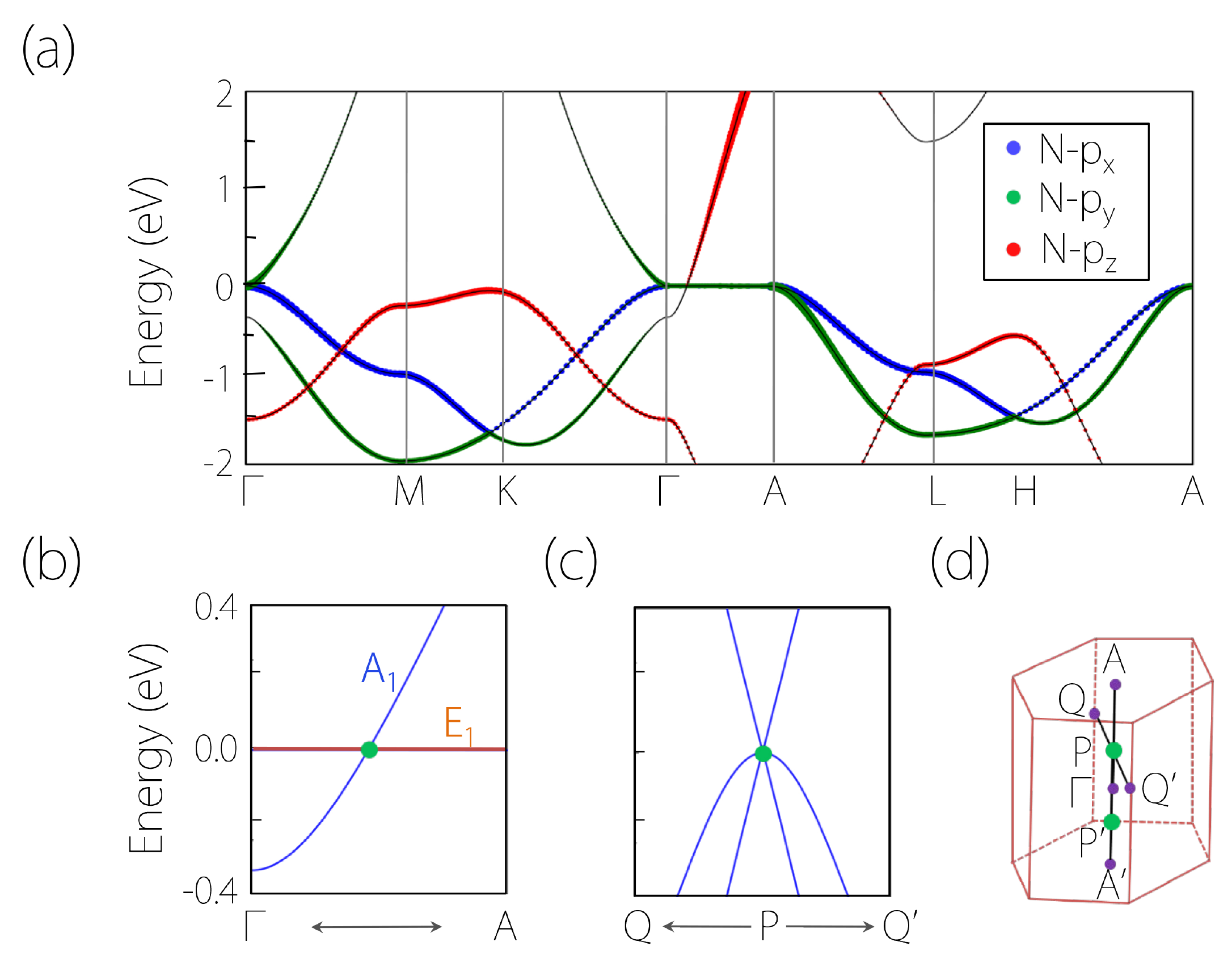}
\caption{(a) The orbital-projected band structure of Li$_2$NaN without SOC. (b) Enlarged view of band structure along the $\Gamma$-A path. (c) Enlarged view of the band structure along the Q-P-Q' path, showing the band dispersion in the (100) direction around the TDNP. (d) The bulk Brillouin zone. P and P¡¯ show positions of TDNPs. Q and Q' are two points in the (100) direction with \emph{k$_z$} set to the TDNP.
}
\label{fig2}
\end{figure}

The orbital-projected band structure for Li$_2$NaN is plotted in Fig.~\ref{fig2}(a). Here, we did not consider SOC in the calculation since the SOC strength is relatively small in Li$_2$NaN (because the material only contains light elements), and we will discuss its effect later. In the band structure, we clearly observe a linear band crossing point on the $\Gamma$-A path at the Fermi level. Actually, the crossing point has a triple degeneracy, because it is formed by the crossing of a non-degenerate band and a doubly-degenerate band. By analyzing the orbital components, we find that the non-degenerate band mainly originates from N-\emph{p$_z$} orbital and the double-degenerate band is mostly contributed by N-\emph{p$_x$} and N-\emph{p$_y$} orbitals. Symmetry analysis shows the two bands respectively belong to irreducible representations \emph{E$_1$} and \emph{A$_1$} of the \emph{C$_{6v}$} symmetry for the $\Gamma$-A path, as indicated in the enlarged band structure in Fig.~\ref{fig2}(b). For the paths away from the $\Gamma$-A, the degeneracy of the bands around the TDNP is fully split. In Fig.~\ref{fig2}(c) we show the band dispersion in the (100) direction with \emph{k$_z$} tuned to the TDNP, and the corresponding \emph{k}-path is displayed in Fig.~\ref{fig2}(d). We can observe a linear band crossing are superimposed with a quadratic band, and the three bands cross at the same point, being similar with most TDNP materials~\cite{19,20,21,22}. Due to the presence of time reversal symmetry, there is a pair of such TDNPs, locating on both sides of the $\Gamma$ point, as illustrated in Fig.~\ref{fig2}(d).

We notice that the TDNPs in Li$_2$NaN have following features. As shown in Figs.~\ref{fig2}(a)-(c), there only exists a single pair of TDNPs in the BZ, and the TDNPs just situate at the Fermi level. The TDNPs are naturally far away from the $\Gamma$ point (at a $\sim$ 1/5 distance of the $\Gamma$-A line). More importantly, the band structure around the TDNPs is quite clean in the sense that near the Fermi energy there is no other extraneous band. Therefore, Li$_2$NaN satisfies all the requirements for an ideal TDNP semimetal.

Besides, we also have following remarks. First, TDNPs can be classified into type-A and type-B based on the connection condition for the pair of TDNPs~\cite{19}. The type-A TDNPs are connected by a single nodal line, while the type-B TDNPs are accompanied by four nodal lines. Here, from Figs.~\ref{fig2}(a) and ~\ref{fig2}(d), we find that the TDNPs in Li$_2$NaN belong to type-A. Second, unlike most TDNPs identified in noncentrosymmetric materials~\cite{17,18,19,20,21,22,24,25,26,27,28,29}, the TDNPs in this work exist in the centrosymmetric system. Such centrosymmetric TDNP semimetal have only been reported previously in TiB$_2$ material~\cite{23}, along with distinct properties (such as novel topological phase transition upon SOC) proposed. So Li$_2$NaN can provide another platform to investigate the unique properties of centrosymmetric TDNP semimetals. Third, around the TDNPs, the band structure manifests a critical-type (between conventional type-I and type-II) band crossing, which is formed by a flat band (double-degenerate) and a dispersive (non-degenerate) band. Similar critical-type band dispersion was previously proposed in Dirac semimetals~\cite{43,44,45} and nodal line semimetals~\cite{46}, while has not been identified in TDNP semimetals yet. Considering that the critical-type Dirac semimetals and nodal line semimetals have shown distinct properties with their type-I (or type-II) counterparts, future exploration of the unique properties for the critical-type TDNPs may be meaningful, during which Li$_2$NaN can serve as an excellent platform.

\begin{figure}
%\scalebox{0.14}{\includegraphics*{Figure3.pdf}}
\includegraphics[width=8cm]{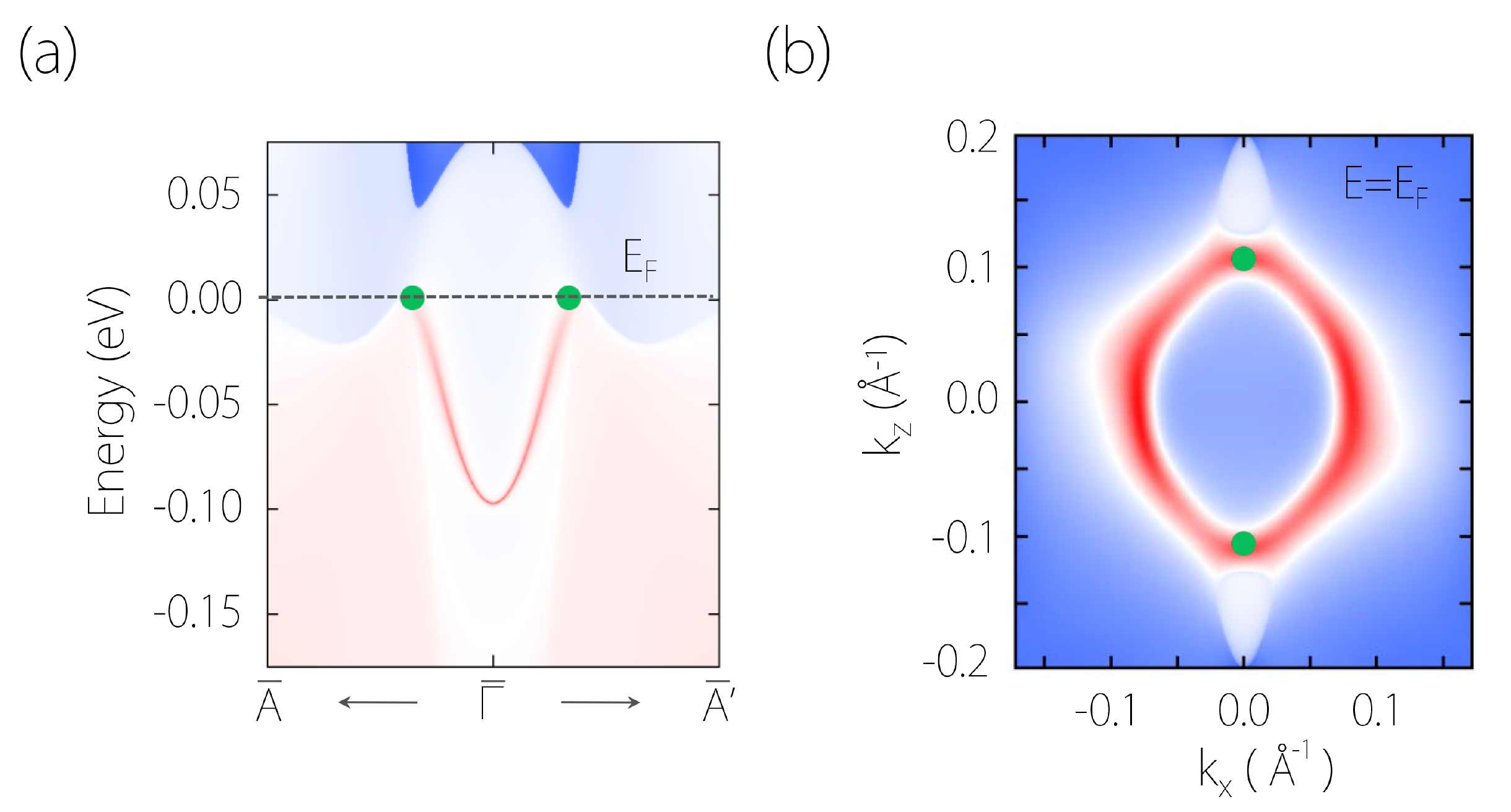}
\caption{(a) Projected spectrum on the (010) surface of Li$_2$NaN. (b) is the corresponding constant energy slice at the Fermi level. In (a) and (b), the green dots mark the position of TDNPs.}
\label{fig3}
\end{figure}

Topological materials characterize with nontrivial surface states. For example, topological insulators show Dirac-cone surface spectrum~\cite{47,48}, nodal line semimetals feature the drumhead~\cite{49,50,51,52,53,54,55,56,57,58,59,60,61,62,63,64,65,66} or cone-like~\cite{23} surface states, Weyl semimetals, Dirac semimetals, and TDNP semimetals usually manifest Fermi-arc surface states~\cite{1,2,3,4,5,6,7,8,9,10,11,12,13,14,17,18,19,20,21,22,23,24,25,26}. For Li$_2$NaN, we show the projected spectrum for the (010) surface and the corresponding constant energy slice at the Fermi level in Figs.~\ref{fig3}(a) and~\ref{fig3}(b), respectively. The exact positions of TDNPs are indicated. We indeed observe the Fermi arc connecting the pair of TDNPs in Li$_2$NaN. These Fermi arc surface states are quite promising to be detected by surface-sensitive probes such as ARPES.

\begin{figure}
%\scalebox{0.2}{\includegraphics*{Figure4.pdf}}
\includegraphics[width=8cm]{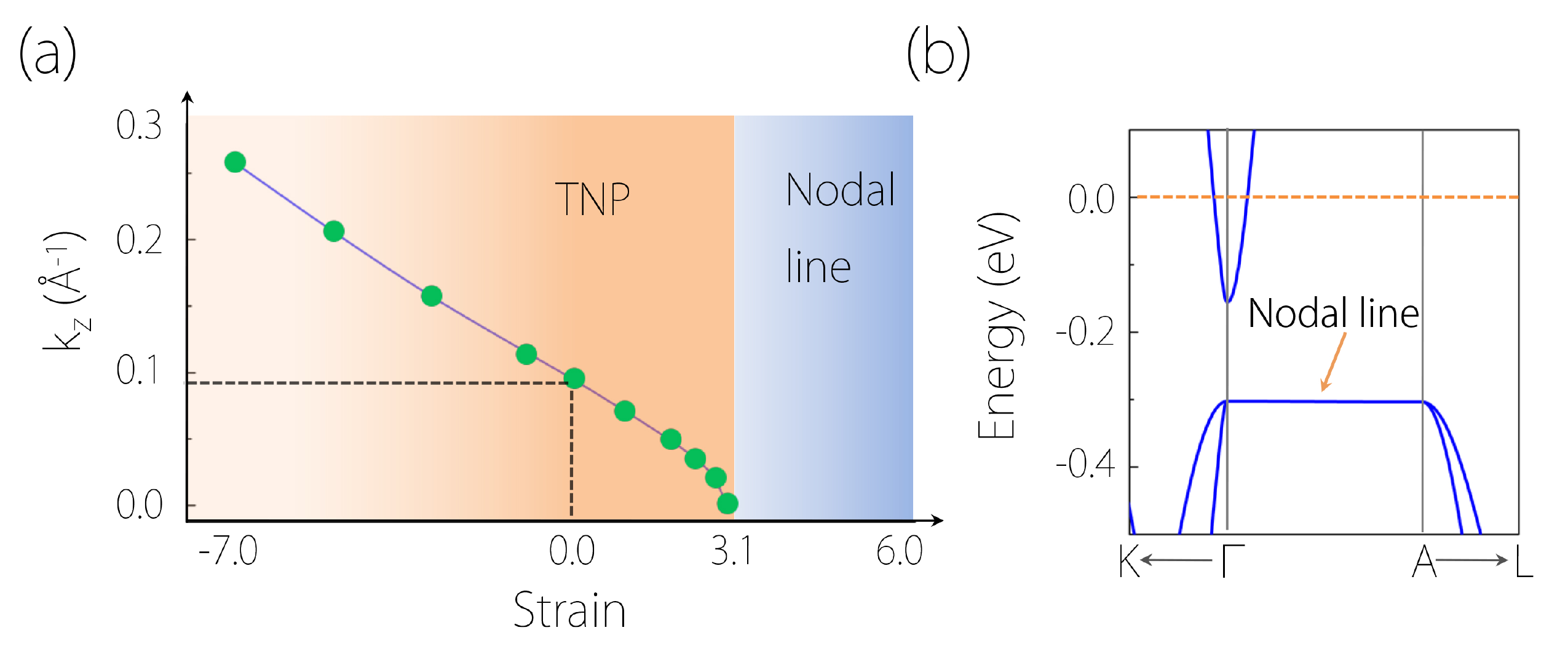}
\caption{(a) Phase diagram of Li$_2$NaN under biaxial strain. The vertical coordinates show the position of TDNPs on the $\Gamma$-A path with respect to (0, 0, \emph{k$_z$}). (b) Band structure of Li$_2$NaN under 5\% tensile strain. The arrow in b points the nodal line on the $\Gamma$-A path.
 }
\label{fig4}
\end{figure}

The TDNPs in Li$_2$NaN are protected by the \emph{C$_{6v}$} symmetry of the $\Gamma$-A line in the absence of SOC. They are robust against weak perturbations with the \emph{C$_{6v}$} symmetry preserved. However, a stronger perturbation may remove the TDNPs by pulling the crossing bands apart, during which the TDNPs annihilate. Such scenario can be realized by strain engineering. Here, we apply the biaxial strain in the \emph{a-b} plane. The phase diagram of Li$_2$NaN under strain is shown in Fig.~\ref{fig4}(a). Under compressive strain, the TDNPs move away from the $\Gamma$ point along the $\Gamma$-A line. While for tensile strain larger than 3.1\%, the TDNPs annihilate and only leave a nodal line on the $\Gamma$-A path, making the system into a nodal line phase. The representative band structure for the nodal line phase is illustrated in Fig.~\ref{fig4}(b). Thus, strain can serve as an effective method to tune the TDNPs in the Li$_2$NaN system.

\begin{figure}
%\scalebox{0.2}{\includegraphics*{Figure5.pdf}}
\includegraphics[width=8cm]{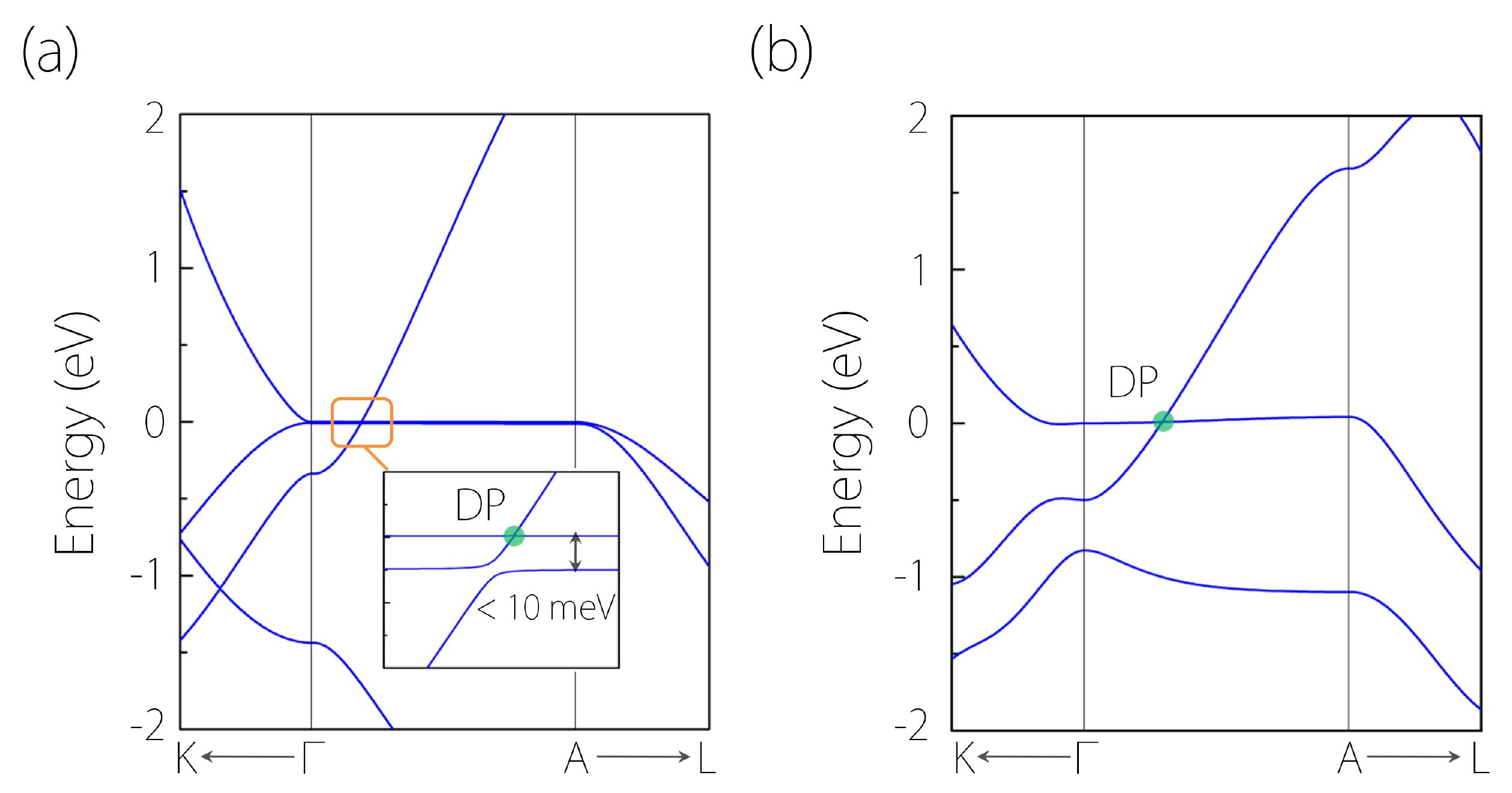}
\caption{Band structures of (a) Li$_2$NaN and (b) Li$_2$NaBi with SOC included. The inset in (a) shows the zoom-in band structure around the band crossing point. In (a) and (b), the green points denote the Dirac point (DP). }
\label{fig5}
\end{figure}

Now we turn to the band structure under SOC. Considering the bands near the Fermi level mainly originate from the N atomic orbitals [see Fig.~\ref{fig2}(a)], the SOC strength is expected to be relatively small since N is a light element. This has been verified by our DFT calculation. Fig.~\ref{fig5}(a) show the band structure of Li$_2$NaN with SOC. We can observe that the band structure is quite similar to that without SOC in Fig.~\ref{fig2}(a). Zooming in the band structure around the TDNP [see the inset of Fig.~\ref{fig5}(a)], we find the flat band is split into two flat bands with a quite small band gap ($< 10$ meV), which ascribes to the Kramers-like degeneracy under SOC. After considering SOC, there also exists a band crossing from a flat band and a dispersive band. The two bands are both doubly-degenerate because of the coexistence of inversion and time reversal symmetries in the system. Thus, this band crossing forms a critical-type Dirac point, as shown in the inset of Fig.~\ref{fig5}(a). Such SOC-induced TDNP-to-Dirac point transition is a typical signature of the centrosymmetic TDNP semimetal~\cite{23}, which does not the preserve in inversion-asymmetric TDNP systems. Here, it is worth noticing that the SOC strength is very small for Li$_2$NaN. At energy scales larger than the SOC-induced band splitting (10 meV), the SOC would have negligible influence on the measured properties, thus the Li$_2$NaN material is quite promising to be well described by a TDNP semimetal. However, if we replace N atom by the much heavier Bi atom, we can obtain an ideal critical-type Dirac semimetal. The band structure of Li$_2$NaBi with SOC is displayed in Fig.~\ref{fig5}(b). Under the strong SOC of Bi atom, the two flat bands are split so much ($> 1$ eV) that only a clean band crossing for a critical-type Dirac point remains near the Fermi level. Therefore, Li$_2$NaN and related materials provide a promising platform for exploring the intriguing properties associated with TDNP semimetals as well as Dirac semimetals.

In conclusion, we predict that an centrosymmetic material Li$_2$NaN is an ideal TDNP semimetal. The material has only a single pair of TDNPs at the Fermi level. The TDNPs naturally situate far away from the $\Gamma$ point, and there is no other extraneous band near the TDNPs. The Fermi arc surface states corresponding to the TDNPs are observed. Even interestingly, the TDNPs are formed by a unique slope of band crossing from a flat band and a dispersive band, leading to the critical-type band structure, which have not been identified in other TDNP semimetals. The TDNPs are protected by the \emph{C$_{6v}$} symmetry in the absence of SOC, and the TDNPs can be efficiently tuned or even annihilated by strain. Unlike most TDNP semimetals identified in noncentrosymmetric materials, the current Li$_2$NaN preserves the inversion symmetry, and features a topological phase transition from the TDNP semimetal to a critical type Dirac point semimetal under the SOC. The SOC strength is found to be quite small, so Li$_2$NaN is well described as a TDNP semimetal. Our work provides an excellent material for exploring the intriguing properties associated with TDNP semimetals in centrosymmetric system.

% This work is supported by the Special Foundation for Theoretical Physics Research Program of China (No. 11747152), Natural Science Foundation of Tianjin City (No.16JCYBJC17200), Research Project for High Level Talent of Hebei Province (No.A2017002020).

%\begin{acknowledgments}
This work is supported by the Special Foundation for Theoretical Physics Research Program of China (No. 11747152), Natural Science Foundation of Tianjin City (No.16JCYBJC17200), Research Project for High Level Talent of Hebei Province (No.A2017002020).
%\end{acknowledgments}

\end{document}